\documentclass[aps,prb,twocolumn,superscriptaddress,showpacs,showkeys]{revtex4-1}
\usepackage{graphicx}
\usepackage{amsmath}
\usepackage{xcolor}

\begin{document}

%\preprint{}

\title{The Effect of Humidity on the Interlayer Interaction of Bi-layer Graphene}


\author{A. Qadir}
\affiliation{College of Information Science and Electronic Engineering, Zhejiang University, Hangzhou 310027, China}
\author{Y. W. Sun}
\email{yiwei.sun@qmul.ac.uk}
\thanks{equal contribution to A. Qadir}
\affiliation{School of Chemical Engineering, University of Birmingham, Birmingham B15 2TT, United Kingdom}
\affiliation{School of Engineering and Materials Science, Queen Mary University of London, London E1 4NS, United Kingdom}
\author{W. Liu}
\affiliation{College of Information Science and Electronic Engineering, Zhejiang University, Hangzhou 310027, China}
\author{P. Goldberg Oppenheimer}
\affiliation{School of Chemical Engineering, University of Birmingham, Birmingham B15 2TT, United Kingdom}
\author{Y. Xu}
\affiliation{College of Information Science and Electronic Engineering, Zhejiang University, Hangzhou 310027, China}
\author{C. J. Humphreys}
\affiliation{School of Engineering and Materials Science, Queen Mary University of London, London E1 4NS, United Kingdom}
\author{D. J. Dunstan}
\affiliation{School of Physics and Astronomy, Queen Mary University of London, London E1 4NS, United Kingdom}

%\thanks{}


\date{\today}

\begin{abstract}
The lubricating ability of graphite largely depends on the environmental humidity, essentially the amount of water in between its layers. In general, intercalated molecules in layered materials modify their extraordinary properties by interacting with the layers. To understand the interaction of intercalated water molecules with graphene layers, we performed Raman measurements on bi-layer graphene at various humidity levels and observed an additional peak close to that of the low-frequency layer breathing mode between two graphene layers. The additional peak is attributed to the vibration between an intercalated water layer and the graphene layers. We further propose that the monolayer coverage of water increases between bilayer graphene with increasing environmental humidity while the interaction between the water layer and graphene layers remains approximately unchanged, until too much water is intercalated to keep the monolayer structure, at just over 50\% relative humidity. Notably, the results suggest that unexpectedly humidity could be an important factor affecting the properties of layered materials, as it significantly modifies the interlayer interaction.
\end{abstract}

%\pacs{62.50.-p, 63.20.-e, 63.20.dk, 63.22.Np}
%\keywords{}

\maketitle

Graphite, the most prominent layered material, has been used as a good lubricant for about 200 years. However, an abnormally high wear rate of graphite brushes in electrical machines aboard aircraft during World War \uppercase\expandafter{\romannumeral 2} was reported. \cite{Przybyszewski68} Later in the 1960s, NASA found that graphite lost its lubrication in space. \cite{Przybyszewski68} It is not the low pressure, but the low humidity, that is primarily responsible for the reduced lubricating ability at high altitude or in space, as a small amount of interlayer water is essential for the easy shear parallel to the graphene hexagonal planes. \cite{Ramadanoff44} A complete understanding of this has not been achieved as it is not yet clear how the interlayer water interacts with the graphene layers.

Layered materials containing different numbers of layers can now be produced, either by exfoliating from the bulk, \cite{Novoselov04} or by stacking one layer onto another. \cite{Suk11} Structures made from different numbers of layers may possess unique properties. \cite{Ferrari06} Intercalated molecules further modify these properties. This vast research area focuses in general on three questions --- how, and how many, intercalating molecules can be introduced between the layers, and what effect they have on pristine layered material. The properties of layered materials largely depend on their interlayer interaction. \cite{Wilson69} It is therefore important to quantify the effect of intercalated molecules on this interaction.

In our work we start with a simple system, bi-layer graphene (the simplest layered structure) with intercalated water (the most common polar molecule), while the amount of water is varied by adjusting the humidity (one of the most common atmospheric conditions).

Much work has been done on water in between graphene oxide layers (GO), as it is hydrophilic --- water easily comes between the layers and attaches to them via non-covalent bonds. \cite{Stankovich07} Early work reported interlayer spacings of 0.63 nm for dry GO \cite{Szabo05} and 1.2 nm for hydrated GO. \cite{deBoer58} The amount of intercalated water can be most accurately obtained by measuring the atomic ratio of carbon to oxygen. \cite{Stankovich06} Nair \textit{et al.} found that permeation of water in GO is unimpeded, and attributed that to a low-friction flow of monolayer water (held by hydrogen bonds) between GO layers. \cite{Nair12} Cerveny \textit{et al.} also reported a water monolayer between GO layers, and it expanded the GO from the interlayer spacing at 0.57 nm to 0.79 nm, when the water content went from 0 to 25 wt\%. \cite{Cerveny10} 25 wt\% interlayer water corresponds to approximately 1 water molecule in every 2 hexagons of carbon. Further expansion to 1.1 nm was observed at 100 \% relative humidity (RH). \cite{Lerf06}  Kim \textit{et al.} measured the diffusivity and dielectric constant of intercalated water in GO layers and found values for both an order of magnitude less than in bulk water, \cite{Kim14} which is further evidence that water has a novel structure when contained between GO layers.

Water is also an essential part of the structure of graphene layers. \cite{Lerf98} Moreover, adsorption of water from the air makes graphene p-type doped. \cite{Ristein06} Yavari \textit{et al.}, by relating the resistivity of graphene to the humidity, found that this doping was stronger when the humidity was increased, but in general it was weak and can be reversed. \cite{Yavari10} Note that this doping was on the graphene-air interface and it is not clear what doping intercalated water might induce.

Techniques used to study water between graphene layers include broadband dielectric spectroscopy (BDS, for conductivity), differential scanning calorimetry (DSC, for heat flow, e.g. calorimetric features of intercalated water were observed at 30 wt\%), X-ray diffraction (for interlayer spacing) and attenuated total reflection geometry in Fourier transform infrared spectroscopy (ATR-FTIR, for the O-H stretching bands). \cite{Cerveny10} We employ Raman spectroscopy to study the interlayer shear mode (CM) and the layer breathing mode (LBM). These modes provide direct measures of the interlayer interaction. Nemanich \textit{et al.} \cite{Nemanich77} first measured the E$_{2g}$ CM of graphite at 42 cm$^{-1}$ and lately Tan \textit{et al.} measured two- to eight-layer graphene and bulk graphite, the frequencies of which fitted well with a linear chain model that considers nearest-neighbour interactions only: \cite{Tan12} 
\begin{equation}
\omega_{NN-i}=\frac{1}{\pi c}\sqrt{\frac{\alpha_{CM}}{\mu}}\sin(\frac{i\pi}{2N}),
\label{eqlc}
\end{equation}
where $\omega_{NN-i}$ is in cm$^{-1}$, \textit{N} is the number of layers, $i = 1, ..., N-1$, denotes the \textit{n}-th mode. \textit{c} is the speed of light in cm s$^{-1}$, $\alpha_{CM}\sim 12.8\times 10^{18}$ N m$^{-3}$ is the interlayer force constant for the CMs, and $\mu=7.6\times 10^{-27}$ kg \AA{}$^{-2}$ is the mass per unit area of monolayer graphene. 

The B$_{2g}$ LBM of bulk graphite is optically inactive and was first measured at 127 cm$^{-1}$ by inelastic neutron scattering. \cite{Alzyab88} It has not been directly probed so far by Raman in Bernal-stacked multi-layer graphene. For twisted multi-layer graphene (we use `twisted' in this paper to be consistent with recent papers on graphene, instead of `turbostratic', from older work on graphite), a similar relation to Eq.\ref{eqlc} applies --- simply replacing the force constant $\alpha_{CM}$ with $\alpha_{LBM}$ ($\sim 115.6\times 10^{18}$ N m$^{-3}$). \cite{Wu15} Sun \textit{et al.} further expanded the interlayer force constant in terms of the interlayer distance as $\alpha=\alpha_{0}(c/c_0)^{6\gamma}$, where $\alpha_{0}$ and $c_0$ are the unperturbed force constant and the interlayer distance respectively, and $\gamma$ is 1.67 for the CM and 2.26 for the LBM. \cite{Sun18}
 
In this work, we loaded a bi-layer graphene sample in a humidity chamber and collected its Raman spectra at different values of the RH. The evolution of the spectra of the G and 2D modes with increased RH  shows that the bilayer graphene behaves as two monolayer graphene sheets with little interaction between the graphene layers, suggesting an increasing amount of water in between the layers. For the LBM, an additional peak was observed, which we interpret as the `breathing' vibrations between an intercalated water layer and graphene layers. The positions of the two LBMs peaks remain nearly unchanged with increasing RH. Therefore, we propose that increasing RH can introduce water into the bilayer graphene and that the water forms a layer structure. The interlayer interaction between graphene layers decreases and the interaction between graphene and the water layer remains approximately constant with increasing RH. At 53 \% RH, the results suggest that intercalated water may no longer maintain the layer structure.

The bilayer graphene samples were prepared on a Si wafer by exfoliation and the fast pick-up technique. \cite{Zomer14} The pick-up was at 100 $^\circ$C and the release was at 160 $^\circ$C. The diameter of the overlapping area of the two graphene layers is about 10 $\mu$m. The sample was placed in a chamber with a humidity sensor, described previously. \cite{Shehzad17} Dry nitrogen flows through the chamber to decrease the humidity while wet nitrogen (bubbled through de-ionised water before entering the chamber) is used to increase it. We performed the room-temperature non-polarised Raman measurements with a Renishaw inVia spectrometer in the backscattering geometry with a confocal microscope. The system has a resolution of 1.3 cm$^{-1}$. We used a 531 nm laser and kept the power on the sample below 5 mW to avoid significant laser heating and consequential softening of the Raman peaks. We should point out that the LBM is only observable under resonance, so the laser wavelength required is determined by the stacking angle of the two graphene layers. The transfer technique we use here gives random angles. With only the fixed 531 nm laser line, we prepared several samples and studied one in which the LBM is in resonance with the laser.

We started the measurements at a low RH of 7\%. At each RH value throughout the experiment we collected two spectra, one in the low wavenumber range for the LBM and CM, the other in the high range for the G and 2D modes. During the collection of a spectrum (20 -- 400 s), slight fluctuations of the RH occurred but did not exceed 1\%. We increased the RH from 7\% to 22\% and then to 37\%. Changes in the spectral profile were observed at 37\%, so we decreased the RH back to 22\%, then increased it again to 30\% and 39\%, and further increased it to 53\%. Losing the signals of the interlayer modes at 53\%, we reduced the RH all the way down to 5\%, and finally brought it back to 16\%.

\begin{figure}
	\includegraphics[width=1\columnwidth]{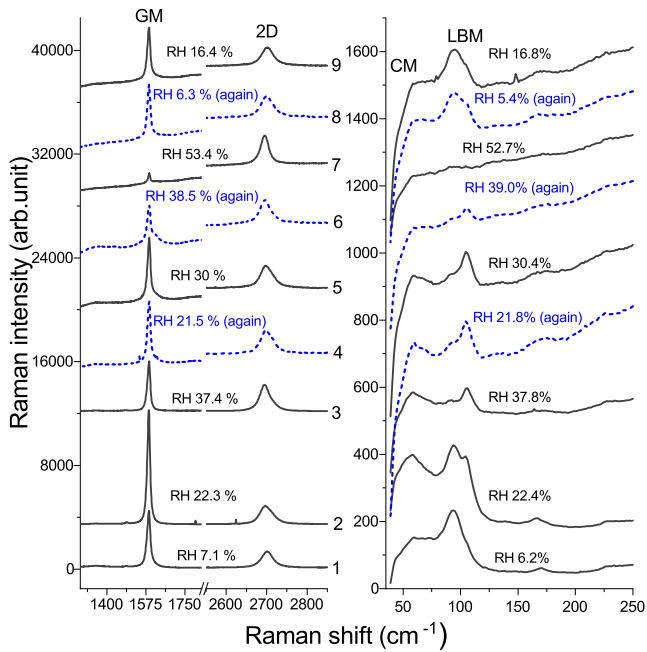}
	\caption{Raman spectra of bilayer graphene at various RH. On the left, the high frequency G and 2D modes are shown, and on the right the low frequency interlayer CM and LBM. All the spectra are vertically shifted for clarity, in the order they were collected, from bottom to top, as numbered. The spectra in black solid lines denote those collected at a higher RH than the previous highest.}
	\label{Graph4}
\end{figure}

The spectra are presented in Fig.\ref{Graph4}. They are vertically shifted for clarity, in the order they were collected, from bottom to top. The spectra in the low and high frequency ranges, collected at the same RH point, are shown at about the same horizontal level (there is a slight difference in the RH values for the spectra on the left and the right of Fig. \ref{Graph4} because of small fluctuations in RH during the collection of spectra, as noted). The key facts are very clear from the spectra, that the intensity ratio of G to 2D significantly decreases at 53 \% RH, and an additional peak close to the LBM clearly appears from 22 \%.

\begin{figure}
	\includegraphics[width=0.85\columnwidth]{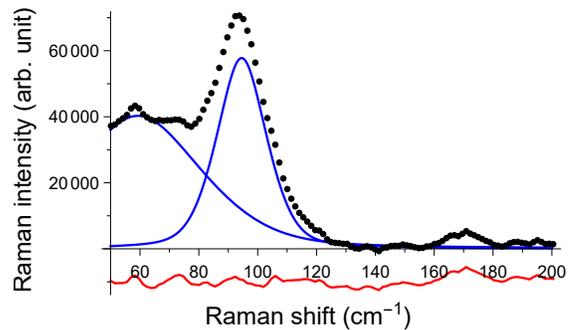}
	\caption{The best fitting of the spectrum collected at 6.2 \% RH and its residuals. The black dots are the spectrum. The background has been subtracted (details in the main text). The blue lines are the two pseudo-Voigts used to fit peaks and the red line is the residuals, vertically downshifted for clarity. We do not fit the peak at $\sim$170 cm$^{-1}$ because it is out of the scope of this paper and the uncertainties from fitting are within the system resolution.}
	\label{eg}
\end{figure}

Spectra were fitted using maximum likelihood estimation. We first subtract the background. For the G and 2D modes, the background of the spectra is relatively flat below the strong Raman peaks. From 21.5 \% RH (blue), we notice a background which then gets stronger. To fit the background, we select a few points on the spectra away from the peak positions and fit them with a 6-term polynomial. For the CM and LBM, although these peaks are intense at low RH, it is challenging to obtain an accurate fit for them as they are close to the cut-off edge of the laser band-reject filter. The background in the range of the CM may consist of contributions from the laser, the edge of the filter, and luminescence. It is difficult and unnecessary to sort every component out. What we do here is to collect two spectra of monolayer graphene in the humidity chamber at 10 \% and 40 \% RH. The line shapes of the two spectra almost overlap, suggesting that the humidity has no effect on the monolayer graphene. The difference between the spectra of monolayer and bilayer graphene is what we are interested in, as it arises from the presence of the second layer. We use the spectra of monolayer graphene at 40 \% RH as background. After subtracting the background, we compare the fittings of various numbers and shapes of peaks by their residuals, and by their values of the Bayesian information criteria (BIC). We show an example of the best fitting for the spectrum collected at 6.2 \% RH, in Fig. \ref{eg}. Here we fit the CM and LBM peaks by two pseudo-Voigt peaks of 57.6 \% Gaussian at 59 and 95 cm$^{-1}$, respectively.

\begin{figure}
\includegraphics[width=1\columnwidth]{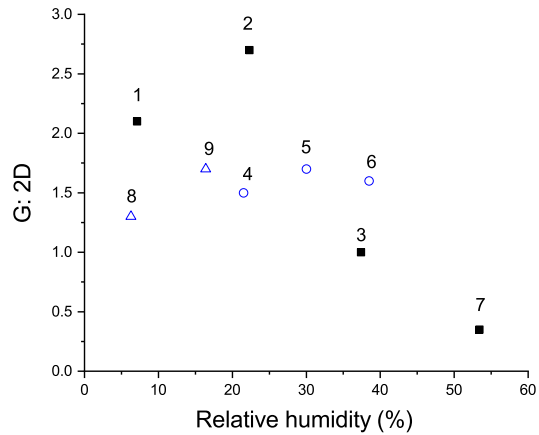}
\caption{The integrated area ratio G:2D of the spectra in Fig. \ref{Graph4} is plotted against RH. Black solid squares denote the spectra collected at a higher RH point than all the previously collected spectra. Blue open circles denote the spectra collected after reducing the RH from  37.4\%, and the blue open triangles are for those collected after reducing the RH from 53.4\%. In both cases, the humidity was decreased and then brought up again. The sequence of the measurements are numbered in accordance with Fig. \ref{Graph4}.}
\label{G2D}
\end{figure}

We first focus on the high frequency range. After subtracting the background, we integrate the spectra (data, not fits) over the G and 2D peaks range, and plot the integrated area ratio of G to 2D in Fig. \ref{G2D}.

We then fit the spectra and find that the GM profile at most RH points consists of two peaks, one at around 1580 cm$^{-1}$, the other at 1590 cm$^{-1}$. There is an additional peak in the range of 1610--1620 cm$^{-1}$ at 38.5\% and 53.4\% RH. These peaks are not resolved in Fig.\ref{Graph4}, but are obtained objectively by the fitting routine described above. No consistent shift of G or 2D peaks with RH is found. Here, it is primarily the intensities of these peaks which is of interest.

The integrated area ratio of G to 2D peaks generally decreases with increased RH. It indicates that the increasing amount of intercalated water weakens the interlayer interaction of the bilayer graphene with increased RH. Using the G to 2D ratio as a qualitative measure of the graphene interlayer interaction, we notice that the bilayer graphene behaves as two monolayers at 52.7 \% where the G: 2D drops much below 1 (typical for monolayer graphene). The process is reversible as the bilayer graphene behaves as a bilayer again when the RH is reduced from 52.7 \% to 5.4 \%.

The results are consistent with the amount of water in between the bi-layer graphene changing its interlayer interaction, and being sensitive to the environmental RH, though there is no definite relation. We would like to point out that bi-layer graphene behaves as two mono-layers at just 53.4\% RH, which can be naturally achieved in the atmosphere of many labs. Additionally, there is an interesting pattern that the ratio slightly goes up and then abruptly drops and it repeats three times --- data points 1$\rightarrow$2$\rightarrow$3, 4$\rightarrow$5$\rightarrow$7, and 8$\rightarrow$9, as shown in Fig. \ref{G2D}.

\begin{figure}
\includegraphics[width=1\columnwidth]{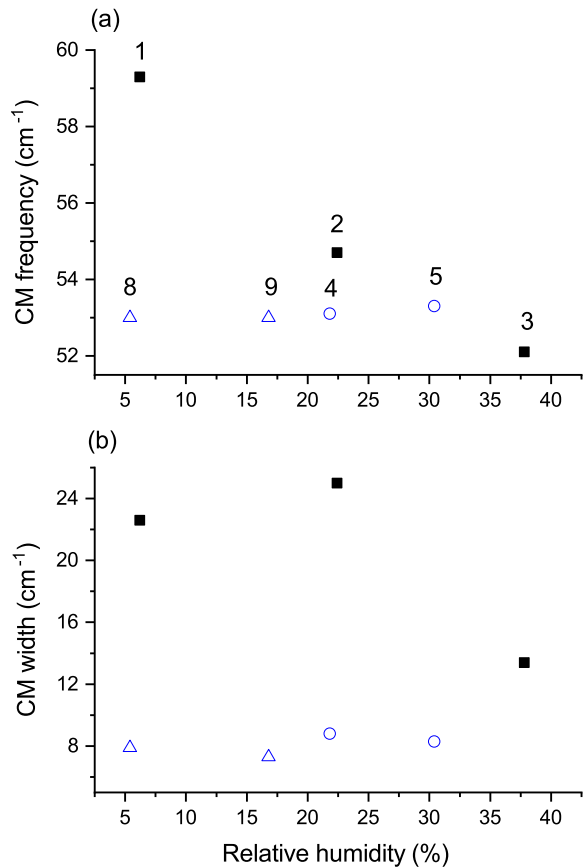}
\caption{(a) Frequency and (b) width of the fitted peak for the CM of bi-layer graphene are plotted against RH. Black solid squares denote the spectra collected at a higher RH point than all the previously collected spectra. Blue open circles denote the spectra collected after reducing the RH from  37.4\%, and the blue open triangles are for those collected after reducing the RH from 53.4\%. In both cases, the humidity was decreased and then brought up again. The sequence of the measurements are labelled in accordance with Fig. \ref{Graph4}. The missing number 6 refers to those of no observable CM signal. The uncertainties in the frequency from fitting are within the system resolution of 1.3 cm$^{-1}$, while the uncertainties in the width are $\sim$ 2 cm$^{-1}$.}
\label{CM}
\end{figure}

We now consider the CM and the LBM. We obtain the frequencies and widths of the peaks from the best fits. Fitted parameters for the CM are plotted against RH in Fig. \ref{CM}. The uncertainties in peak positions from fitting are within the system resolution of 1.3 cm$^{-1}$, while the uncertainties in the width are $\sim$ 2 cm$^{-1}$, at various RH values. The sequence of the measurement is numbered in accordance with the G and 2D modes in Fig. \ref{G2D}. No CM is observed at 38.5 \% as well as 52.7 \% RH. The positions of the CM at various RH values are close, except the very first point at 6.2 \%. No consistent change of width is found.

\begin{figure}
\includegraphics[width=1\columnwidth]{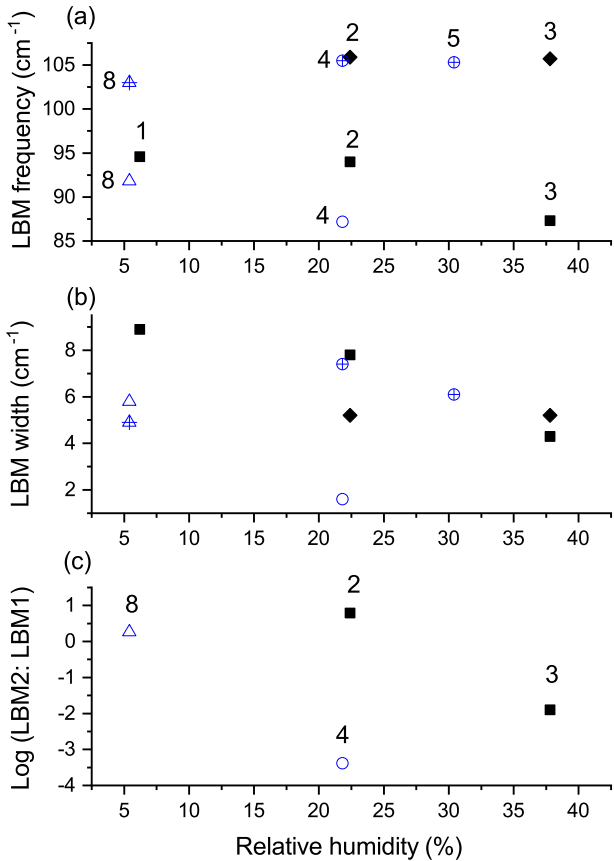}
\caption{(a) Frequency and (b) width of the LBM of bi-layer graphene are plotted against RH. At an RH where there are two LBMs, (c) the integrated area ratio of the LBM from higher to lower frequency is plotted against RH. Black solid squares and diamonds (for the LBM of higher frequency) denote the spectra collected at a higher RH point than all the previously collected spectra. Blue open circles denote the spectra collected after reducing the RH from  37.4\%, and the blue open triangles are for those collected after reducing the RH from 53.4\%. In both cases, the humidity was decreased and then brought up again. The symbols with a cross are for the LBM of higher frequency. The uncertainties in the frequency and width from fitting are within the system resolution of 1.3 cm$^{-1}$. The sequence of the measurements are numbered in accordance with Fig. \ref{Graph4}. The missing numbers refer to those of no clear LBM signal.}
\label{LBM}
\end{figure}

The additional LBM peak is the main finding of this work. In Fig. \ref{LBM} we plot the peak positions and widths of the LBM, and the integrated-area ratio of the higher LBM to the lower,  against RH.

The additional peak appears from the second RH point at 22.4\%. We attribute this peak to the vibration between the intercalated water layer and graphene layers, from the following considerations. First, the decreasing G to 2D ratio with increasing RH indicates the weakened interaction between two graphene layers. We therefore expected a redshift of the LBM. Second, the frequency of the LBM of bi-layer graphene does not change much after water intercalation from 22.4 \% RH. This suggests that the bi-layer graphene was only partially filled by water. Third, the additional peak is at higher frequency than the original LBM of bi-layer graphene, indicating that it is not from vibrations of two graphene layers but perhaps involves something of greater mass. Also, the nearly unchanged position of this additional peak at various RH values indicates that it corresponds to an interaction between layers that does not change much with increasing amounts of intercalated water. On both counts, it can be attributed to the water-filled regions.  Fourth, adsorption of water on the graphene makes it p-type doped. It is known that doping shifts the GM, but no evidence shows that doping shifts the LBM. With an increasing amount of water on both sides of the top graphene layer, one would expect an increasing level of doping. Again the approximately unchanged position of the additional peak at various RH values suggests that it is not just a doping-modified LBM of graphene layers. Fifth, the reported monolayer of water between GO layers \cite{Nair12} seems relevant here. The additional peak can be considered as the vibration between graphene and the water layer. With increasing amounts of intercalated water, the interaction between graphene and the water layer remains nearly unchanged, while the coverage of the water layer increases, until the amount of water is too much to maintain the layer structure at 52.7 \%, where all the interlayer modes disappear. Finally, we apply a linear chain model to obtain a force constant for this additional layer breathing mode. The chain consists of graphene, water and graphene. We consider the nearest-neighbour interaction only. We assume that there is one water molecule in the area of a carbon hexagon, in terms of the density of the water layer, as it has to be condensed enough to form a layer, and held by hydrogen bonds following Nair et al. \cite{Nair12} (if we consider the bonding of water in-between two graphene layers to be the same as between GO). We insert $i=1$, $N=3$, the reduced mass (of water and carbon) $m=3.26\times 10^{-27}$ kg \AA{}$^{-2}$, and the measured LBM frequency to Eq. \ref{eqlc} and obtain the force constant $\alpha_{Gr-water}=37.9\times 10^{18}$ N m$^{-3}$, about 1/3 of the force constant between graphene layers. This is a reasonable value for the force constant. All of the LBM peaks are very narrow (width below $\sim$10 cm$^{-1}$), and there is no abrupt change in the width of fitted peaks, further validating the reliability of the presented results, on which the above discussion is based. The integrated area ratio of the two LBM changes with RH but no clear relation is observed.

In summary, we measured the high frequency G and 2D modes, and the low frequency interlayer CM and LBM of bi-layer graphene at various humidity levels. With increasing RH, we observe a decreasing intensity ratio of the G to 2D mode, a downshift of the CM and notably an additional LBM. We conclude that intercalated water molecules form a layer, the interaction of which with the graphene layers is about 1/3 of that between pristine graphene layers. With increasing amounts of water, the interlayer interaction of the bilayer graphene is weakened, while the interaction between graphene and the water layer remains nearly unchanged, until too much water is intercalated to maintain the layer structure at over 50 \% RH. Water molecules can be introduced in between graphene layers at relatively low levels of the humidity, and by increasing the humidity the bi-layer graphene behaves as two monolayers, at just over 50 \% RH. This suggests that atmospheric humidity could be a crucial parameter in many laboratories affecting the results of graphene-related research and applications.

The authors are grateful for valuable comments by Prof. P. H. Tan from the Chinese Academy of Science, and by Prof. A. J. Drew from Queen Mary University of London.

\bibliography{apssamp1}

\end{document}